\documentclass[%
 reprint,
superscriptaddress,
 amsmath,amssymb,
 aps,
]{revtex4-2}

\usepackage{graphicx}
\usepackage{dcolumn}
\usepackage{bm}
\usepackage{ulem}
\usepackage{xcolor}

\begin{document}

\preprint{APS/123-QED}

\title{Air entrainment by an inclined smooth water jet}

\author{Théophile Gaichies }
\affiliation{Universit\'e Paris-Saclay, CNRS, Laboratoire de Physique des Solides, 91405 Orsay, France}
\affiliation{Carlos III University of Madrid, Department of Thermal and Fluids
Engineering, Leganés, 28911, Spain}
\author{Arnaud Antkowiak}
\affiliation{Sorbonne Université, CNRS, Institut Jean le Rond $\partial$’Alembert, F-75005 Paris, France}
\author{Anniina Salonen}
\affiliation{Universit\'e Paris-Saclay, CNRS, Laboratoire de Physique des Solides, 91405 Orsay, France}
\affiliation{Soft Matter Sciences and Engineering, ESPCI Paris, PSL University, CNRS, Sorbonne Université, 75005 Paris, France}
\author{Emmanuelle Rio}
\affiliation{Universit\'e Paris-Saclay, CNRS, Laboratoire de Physique des Solides, 91405 Orsay, France}

\date{\today}

\begin{abstract}
Air entrainment can occur when a water jet impacts a water/air interface, a process central in various real systems, ranging from dam spills to breaking waves. Despite its prevalence, a comprehensive description of the mechanism controlling bubble size distribution remains elusive.
Here, we establish a link between the geometry and the dynamics of the cavity observed when an inclined impinging jet impacts a water interface and the resulting bubble cloud.
We show that the bubbles result from the destabilization of the wavefield developing at the interface of the cavity.
The origin of this wave field is the creation of a shear layer, due to the asymmetric detachment of the flow field from the interface. 
\end{abstract}

\maketitle

Air entrainment by impinging jets has been the subject of a profusion of investigations \cite{kiger2012air,lorenceau2004air,bin1993gas}, serving as a model system for complex natural flows such as breaking waves \cite{koga1982bubble} and dam spills \cite{guyot2020penetration}.
However, while the entrainment mechanism is well-characterized for viscous liquids \cite{jeong1992free,eggers2001air,lorenceau2003fracture,lorenceau2004air} and despite the fact that water is ubiquitous in practical and industrial applications, it remains poorly understood for low-viscosity fluids like water \cite{kiger2012air}. 
Previous studies have shown that surface irregularities trigger entrainment by vertical jets \cite{zhu2000mechanism,redor2025air}. 
Yet, for smooth jets of millimetric radius, air entrainment requires symmetry breaking, such as underwater vortice formation \cite{mckeogh1981air,kiger2012air}, jet translation \cite{chirichella2002incipient}, or inclination \cite{miwa2019experimental,koga1982bubble}. 
In particular, jet inclination drastically alters the interface topology \cite{gaichies2026shapeinterfacehitoblique} and is known to reduce the velocity threshold for air entrainment \cite{koga1982bubble,detsch1990critical}.
The origin of this diminution is however unclear. 

In this Letter, we elucidate the physical mechanism of air entrainment for a smooth, inclined water jet. 
By combining high-speed imaging with direct numerical simulations, we reveal that the process is driven by a shear instability. We show that the flow separation in the acute part creates a free shear layer in the liquid phase. 
The destabilization of this layer generates vortices, exciting waves on the cavity's surface that subsequently break up into bubbles. 
This work establishes a link between the vorticity generated at the impact site and the statistics of the entrained bubble cloud, offering a comprehensive view of aeration in asymmetric continuous impact flows. 

\begin{figure}[h!]
    \centering
    \includegraphics[width=0.46\textwidth]{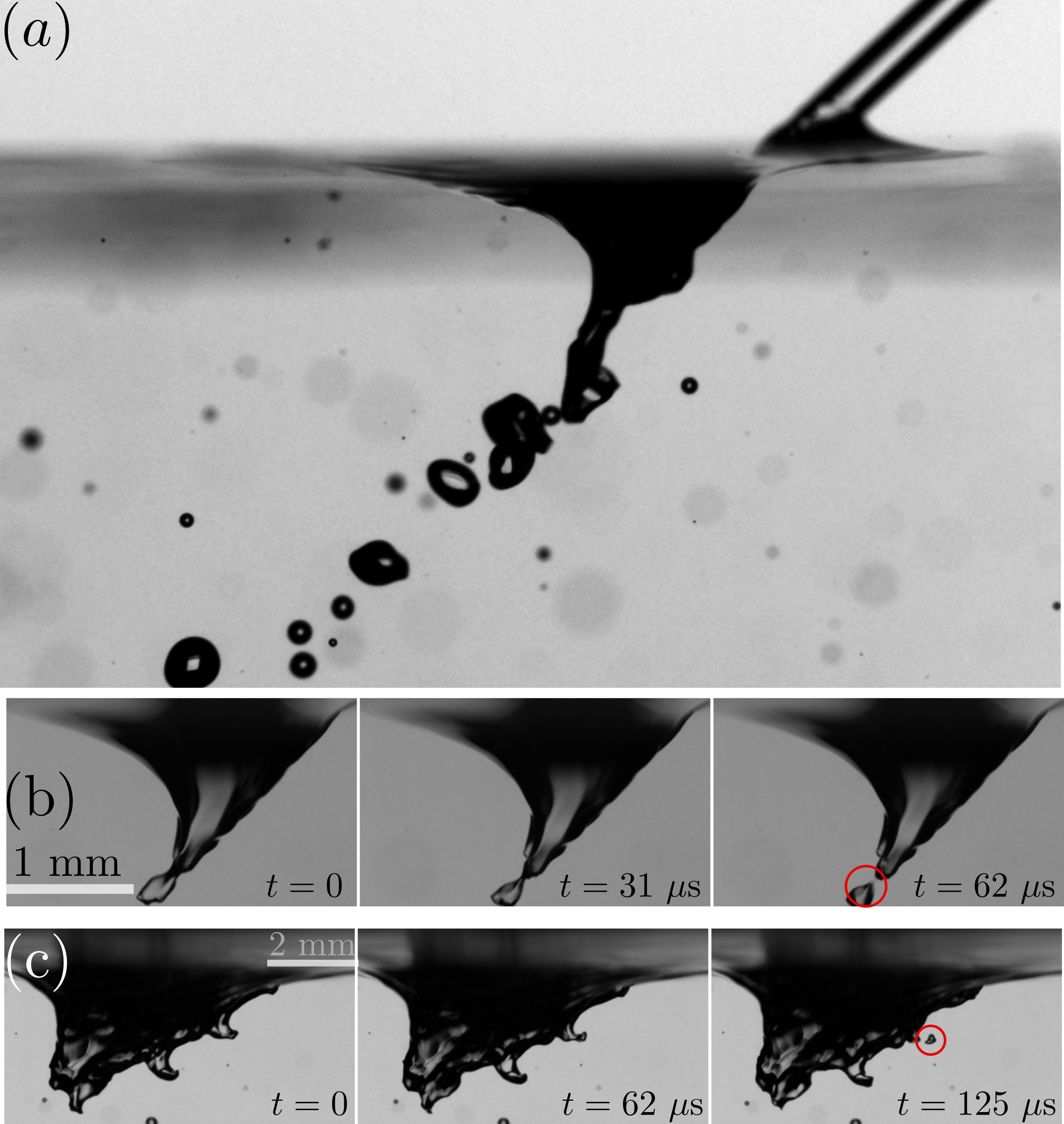}
    \caption{(a) Image of the bubble cloud produced by an inclined jet with $R = 0.29\rm\ mm$, $V = 3.1 \rm \ m.s^{-1}$, and $\alpha = 43^{\circ}$. (b) Image sequence of an air entraining event at the tip of a cavity formed by a jet with $R = 0.19\rm\ mm$, $V = 4.1 \rm \ m.s^{-1}$, $ \alpha = 42^\circ$. (c) Image sequence of air entraining events along the side of a cavity formed by a jet with $R = 0.4\rm\ mm$, $V = 2.6 \rm \ m.s^{-1}$ , $ \alpha = 32^\circ$. The red circles show the site where the bubbles are formed.}
    \label{fig:intro}
\end{figure}

A typical air cavity leading to bubble generation is shown in Fig. \ref{fig:intro}(a). The jets creating such cavities are generated by connecting a liquid reservoir, a 500 mL glass bottle (Schott), to a pressure controller (Elveflow ob1 mk3). 
The overpressure imposed in the reservoir drives the liquid (Milli-Q water) through fluoropolymer tubing (Versilon FEP Saint-Gobain), resulting in a laminar jet upon exiting a stainless steel nozzle (Nordson EFD Optimum). 
The materials have been carefully chosen to avoid any contamination of the liquid.
By changing the nozzle and the pressure, we are able to study jets of radius $R \in [0.12,0.4] \rm \ mm$  and speed $V \in [1.4,5.3]\ \rm m.s^{-1}$. 
The nozzle is mounted on a rotating axis, allowing us to impose an angle between the jet and the horizontal, $\alpha \in [23,48]^\circ$. 
This inclined jet impacts a bath of the same liquid contained in a 10x5x5 cm tank, indented at the top to maintain a steady water level during the experiment. 
To image the interface, we use a backlight (Phlox led LLUB 50x50) with a fast camera (FASTCAM NOVA S9), connected to a macro lens (Laowa 2x or 2x5).

\begin{figure}[h!]
    \centering
    \includegraphics[width=0.46\textwidth]{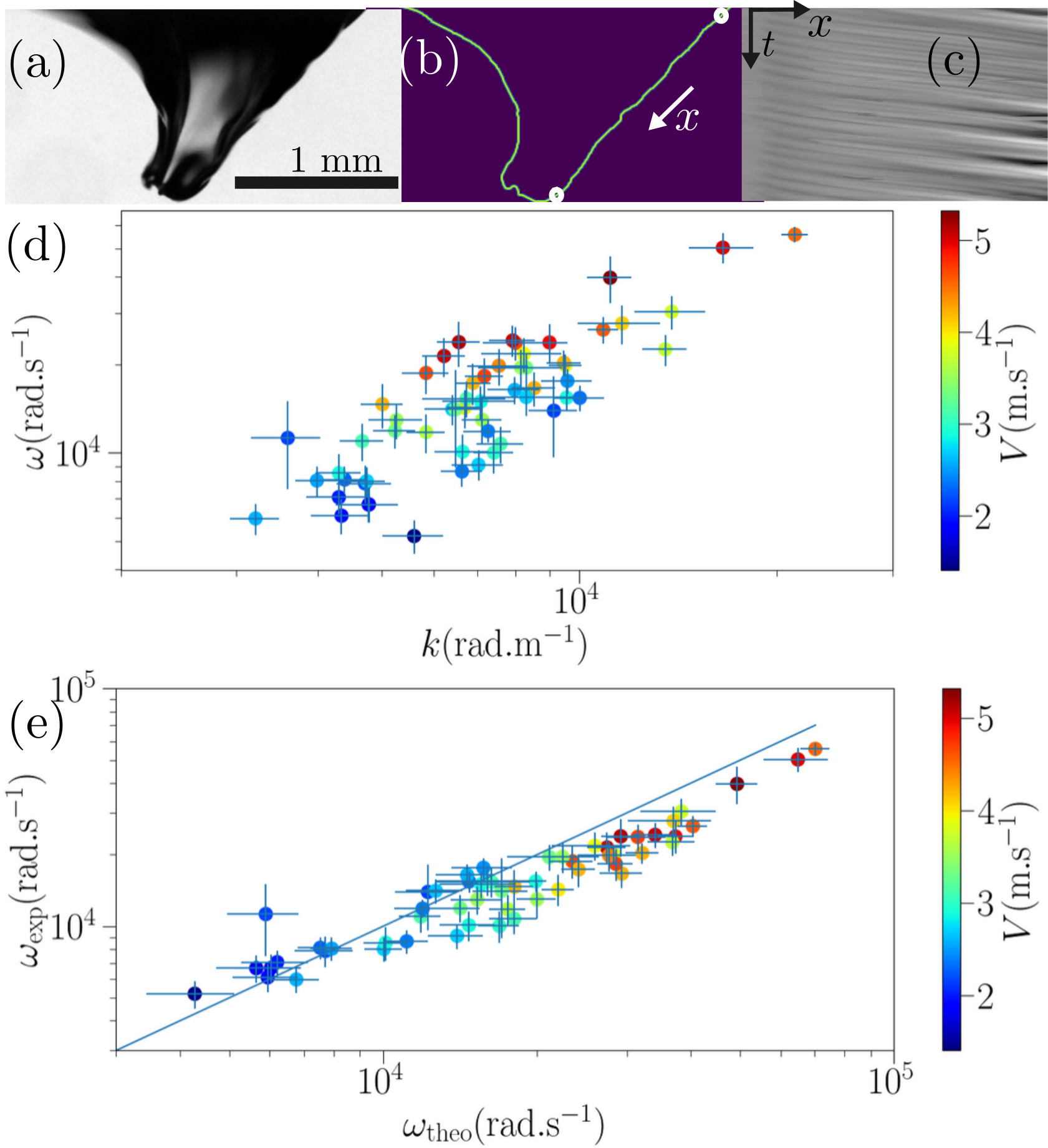}
    \caption{(a) Experimental image of the cavity formed by a jet with $R = 0.19\rm\ mm$, $V = 4.1 \rm \ m.s^{-1}$ , $ \alpha = 42^\circ$. (b) Contour of the cavity extracted from (a). For further processing, only the part situated between the two white dots is kept. (c) Image obtained by stacking successive profiles represented by single lines, where the elevation is encoded in gray level. (d) Measurements of the angular frequency plotted against the associated wavenumber. (e) Comparison between the measured angular frequency of the wavefield with the prediction of the angular frequency computed from the measured wavenumber with the dispersion relation of a shear driven instability. The plots with color encoding the radius or the angle of the jet can be found in the supplemental materials (For (d) Fig. S1 and for (e) Fig. S2 \cite{suppmat}).}
    \label{fig:wave}
\end{figure}
In Fig.\ref{fig:intro}(a), we can see an inclined jet creating a cavity in front of its impact site, which can deform and create bubbles (see movie V1 \cite{suppmat}). 
The existence and the shape of the cavity are the topic of another article \cite{gaichies2026shapeinterfacehitoblique}. 
Here, we will focus on the processes that perturb the cavity's shape and lead to air entrainment. 
To observe the creation of the bubbles, it is necessary to image the cavity at a higher framerate. 
Typical image sequences capturing bubble formation are displayed in Fig.\ref{fig:intro}(b) and (c).
In these figures, propagating surface waves can be observed, riding along the right part of the cavity. 
Their growth can result in rupture of the interface and bubble entrainment, mainly at the tip of the cavity when $\alpha \gtrsim  40^\circ$ (Fig.\ref{fig:intro}(b), movie V2 \cite{suppmat}) , or along the side of the cavity when $\alpha \lesssim 40^\circ$ (Fig.\ref{fig:intro}(c), movie V3 \cite{suppmat}). In both cases, the waves are the main driver behind the creation of the bubbles. This prompts us to investigate in more detail the wavefield developing on the side of the cavity, and to try to link its characteristics with the bubbles created by the cavity. To do so, we perform experiments filming the cavity at 64 000 frames per second. We keep 12 stacks of 500 consecutive images distributed evenly over 2 seconds of experiments. For the same set of parameters, we also image the bubble cloud created by the cavity at 9000 images per second for half a second. The results of this second batch of experiments will be discussed in Fig. \ref{fig:bubble}.

In order to characterize the wavefield, we first extract the contour of the cavity. Fig.~\ref{fig:wave}(a) and \ref{fig:wave}(b) report a typical cavity snapshot and the corresponding extracted contour. The rightmost part of the contour, delimited by dots in the picture, is further post-processed: after rotation, this deformation profile is turned into a single pixel line with the gray level encoding the local height. The lines extracted over time are then stacked, and a final matrix is obtained such as the one in Fig.\ref{fig:wave}(c). For each set of parameters, each stack of 500 consecutive images is transformed into a spatio-temporal matrix. 
From the Fourier transform of these matrices, we retain all peaks exceeding 25$\%$ of the maximum intensity. We then extract the centroids from 5×5 pixel neighborhoods centered on these peaks. Finally, a weighted average based on normalized intensities is calculated across all the centroids extracted from the 12 matrices to yield single values for $\omega$ and $k$. 
The results of these measurements are displayed in Fig.\ref{fig:wave}(d). For a given $k$, we can observe that $\omega$ increases with the flow velocity $V$. This behavior is consistent with a shear-driven instability.

The typical Kelvin-Helmholtz dispersion relation for a shear instability between two fluids having velocities $U_1,U_2$, and densities $\rho_1,\rho_2$ is  \cite{rabaud2020kelvin}: 
$$ \rho_1 (U_1 - \frac{w}{k})^2 + \rho_2 (U_2 - \frac{w}{k})^2 = (\rho_1-\rho_2)g/k + \gamma k,$$
where $g$ is the gravitational acceleration and $\gamma$ the surface tension. In the case of the cavity, the two fluids are water and air, so  $\rho_1 =\rho_{\rm water} \gg \rho_2$. The air is at rest initially, so we assume $U_1 \sim V \gg U_2$. As the typical wavelength of the instability is small compared to the capillary length, we also neglect gravity. Finally, we obtain the simplified dispersion relation (choosing a negative correction of the surface tension) $\omega(k) = Vk - (\gamma k^3/\rho_{\rm water})^{1/2}$. To verify that this model is compatible with our experimental results, we put the measured $k_{\rm exp}$ and $V$ in the simplified dispersion relation to compute the predicted angular frequency $\omega_{\rm theo}$. In Fig.\ref{fig:wave}(e), this angular frequency is compared with the one measured experimentally, and we can observe that the agreement is satisfactory, thereby confirming the shear-driven nature of the instability. 

\begin{figure}[h!]
    \centering
    \includegraphics[width=0.46\textwidth]{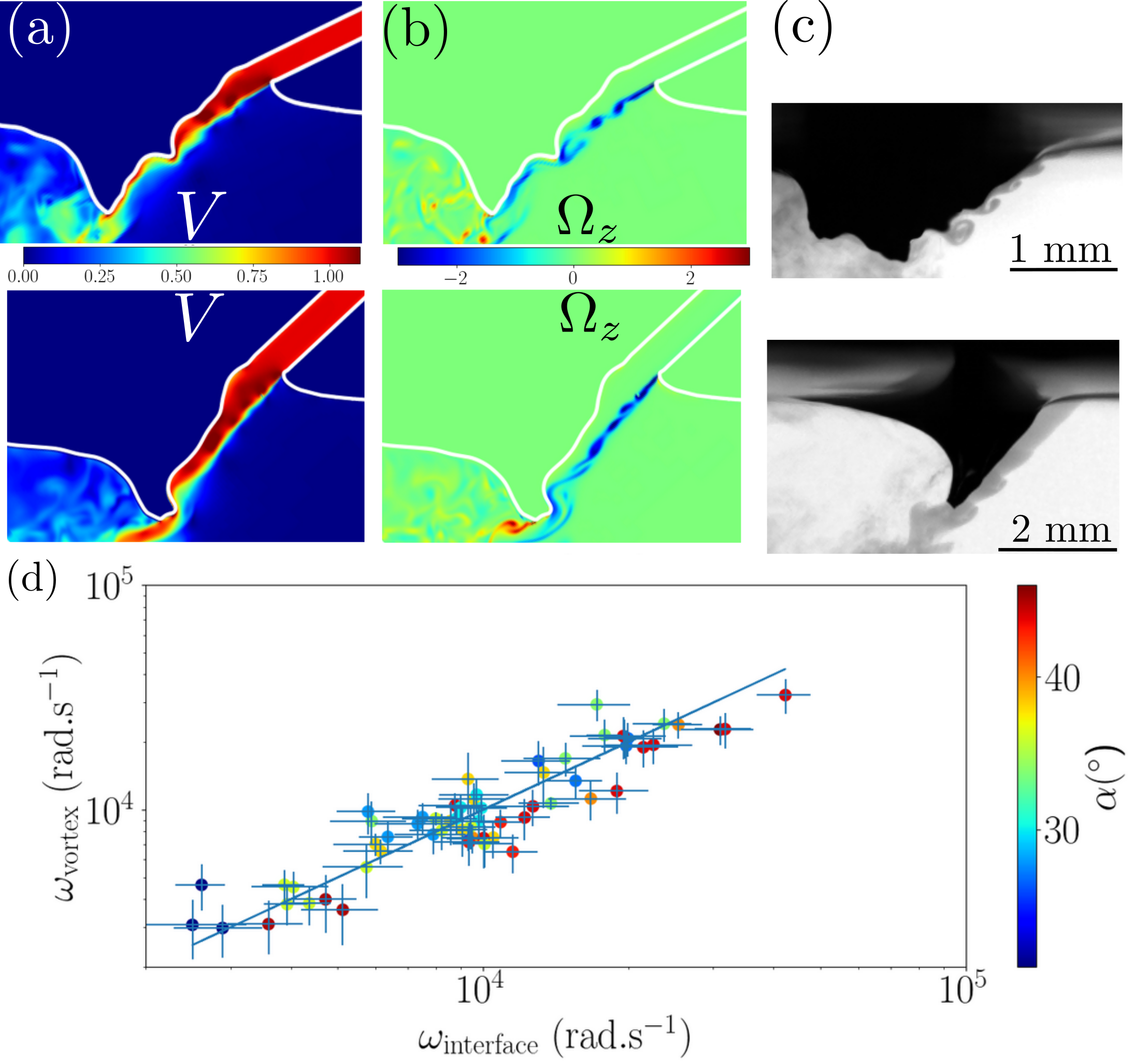}
    \caption{(a) and (b) Velocity (respectively vorticity along the vertical axis) field in the symmetry plane of the system for $\alpha = 26^\circ$ (top) and $\alpha = 43^\circ$. The velocity (resp. vorticity) is normalized by the injection speed of the jet (resp. by the injection speed divided by the radius of the jet). (c) Experimental images of the cavity formed with a stained jet containing 0.5 wt\% of indigo carmine, for a jet with parameters $R = 0.12\rm\ mm$, $V = 3.4 \rm \ m.s^{-1}$, $ \alpha = 27^\circ$ (top), and $V = 3.1 \rm \ m.s^{-1}$, $ \alpha = 43^\circ$ (bottom). (d) Comparison of the experimental angular frequency of the waves in the shear layer and that of the air-water interface on the cavity. The comparison with color encoding the radius or the velocity of the jet can be found in the supplemental materials (Fig. S3) \cite{suppmat}.}
    \label{fig:vortexp}
\end{figure}

To understand the selection of the pulsation, we investigate the flow along the cavity. As experimental measurements near the impact zone are hindered by optical distortion caused by the interface curvature (see Fig.\ref{fig:intro}(a)), we perform direct numerical simulations of the flow with Basilisk, an open source software that has been extensively validated for multiphase flow and air entrainment \cite{Mostert2022,riviere2022capillary,popinet2009accurate}. The physical parameters of the simulations are chosen to match the experimental ones (the details of these simulations can be found in another article \cite{gaichies2026shapeinterfacehitoblique}). We can then obtain the velocity field in the plane of symmetry, as represented for $\alpha= 26^\circ$ and $\alpha= 43^\circ$ in Fig.\ref{fig:vortexp}(a). 
We observe that there is a flow separation from the interface in the acute part, while the flow follows the interface of the cavity in the obtuse part. Because of this separation in the acute part, a shear layer is created which generates vorticity $\Omega_z$ along the axis perpendicular to the plane of symmetry (Fig.\ref{fig:vortexp}(b)). This layer can then destabilize and form vortices. To confirm these numerical observations, we perform a new set of experiments adding indigo carmine in the pressure reservoir to produce a stained jet. As can be seen in Fig.\ref{fig:vortexp}(c), the formation of vortices in the shear layer is also observed. Furthermore, we can observe that the vortices are generated upstream compared to the waves, and that their frequency appears comparable to that of the waves (see movie V4 and V5 \cite{suppmat}).

To confirm this observation, we measure the frequencies of the vortices and the waves.
The angular frequency of the wavefield $\omega_{\rm interface}$ is measured with the previously described method. To measure the angular frequency of the vortex generation $\omega_{\rm vortex}$ we do a 1D Fourier transform along a line perpendicular to the dyed jet, situated before the first vortex roll up. The two frequencies are represented in Fig.\ref{fig:vortexp}(d), which suggests that, to leading order, $\omega_{\rm vortex}\sim \omega_{\rm interface}$. A prediction for the observed selected $\omega_{\rm vortex}$, should then hold for the angular frequency of the selected $\omega_{\rm interface}$. 

\begin{figure}[h!]
    \centering
    \includegraphics[width=0.46\textwidth]{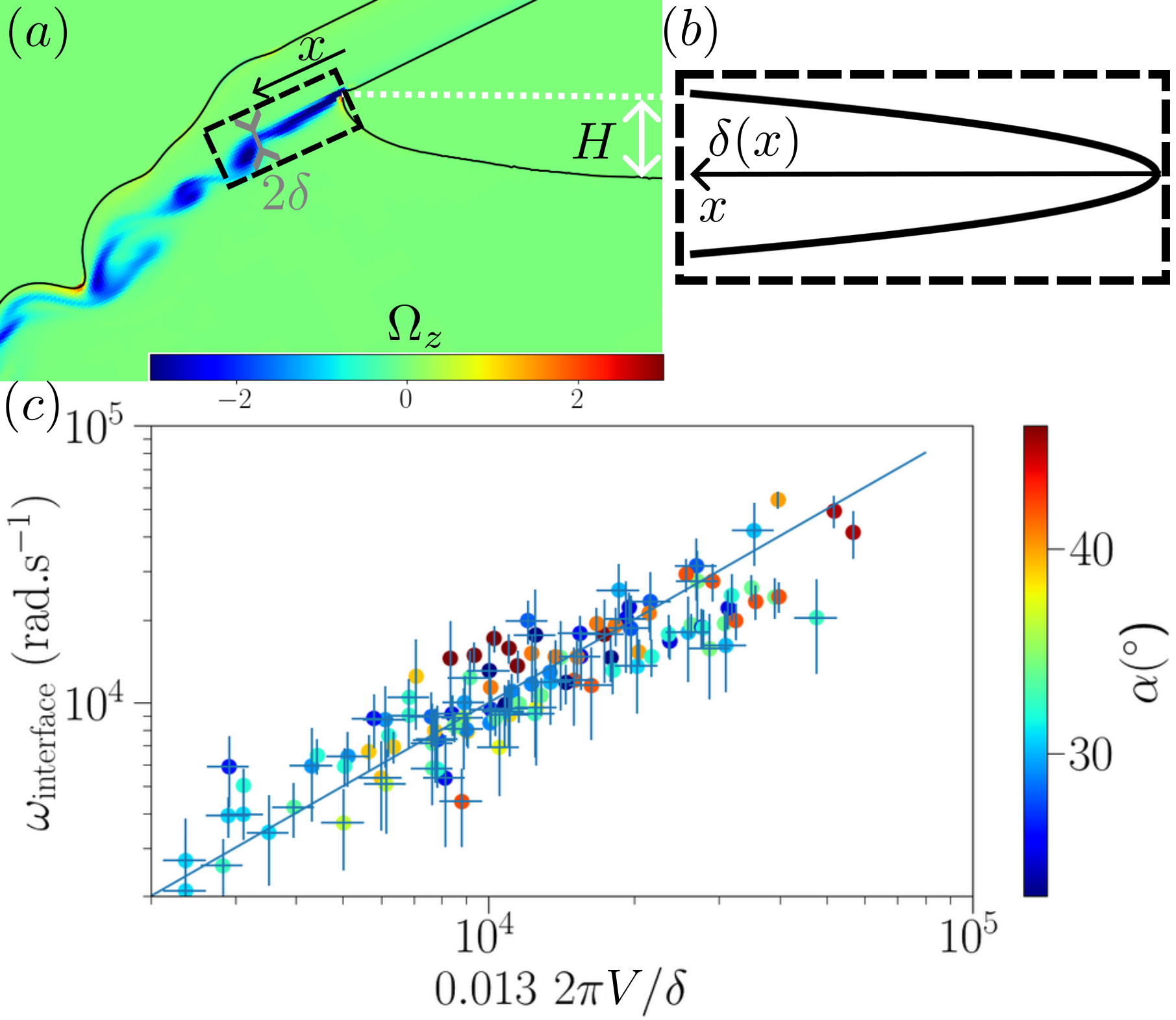}
    \caption{ (a) Sketch of the vorticity and the boundary layer formed during the jet impact. (b) Sketch of the typical laminar wake evolution used to model the shear layer formed by the flow separation in the meniscus. (c) Comparison between the measured angular frequencies propagating on the cavity and the prediction based on equation \ref{eq:final}. The blue line represents the $y=x$ curve. The comparison with color encoding the radius or the velocity of the jet can be found in the supplemental materials (Fig. S4) \cite{suppmat}.}
    \label{fig:vortheo}
\end{figure}

\begin{figure}[h]
    \centering
    \includegraphics[width=0.46\textwidth]{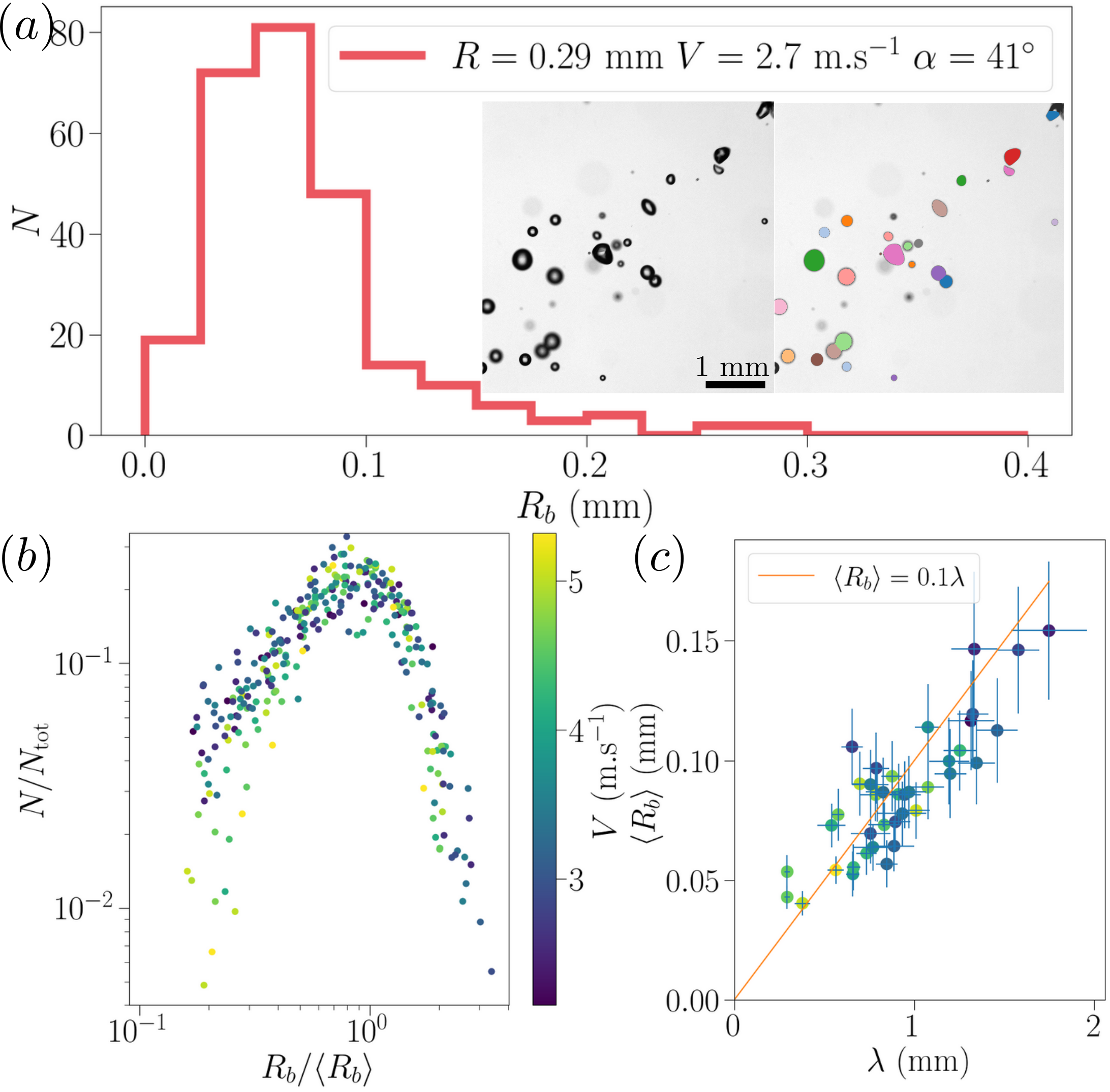}
    \caption{(a) Histogram of the bubble size distribution obtained after the tracking procedure. Inset shows one picture of the image stack before and after segmentation. (b) 37 bubble size distributions for various jet parameters, normalized by the number of bubbles formed during the experiment and the average size of the bubbles. (c) Average size of the bubbles formed during the experiment against the typical wavelength of the instability. }
    \label{fig:bubble}
\end{figure}

The destabilization of free shear layers has been extensively studied in the literature  \cite{ho1984perturbed}. In particular, it has been shown that for planar free shear layer, the primary vortex shedding frequency is associated with a critical Strouhal number $\mathrm{St} = \omega_{\rm vortex}\delta/2\pi V\sim0.032$ \footnote{See Fig.2(a) of the review article  \cite{ho1984perturbed}}, where $\delta$ is the thickness of the shear layer at the point of vortex shedding. It follows that $\omega_{\rm interface} \sim 2\pi V /\delta$. We then need to evaluate $\delta$. As sketched in Fig.\ref{fig:vortheo}(a), the vortex shedding takes place typically at the altitude of the interface, at a distance $H$ below the flow separation point. As shown in \cite{gaichies2026shapeinterfacehitoblique}, this typical height can be written 
\begin{equation}
    H(\alpha)\sim R \ln\left(\frac{1+\cos(\alpha)}{1-\cos(\alpha)}\right).
\end{equation} 
We assume that the  shear layer experiences viscous diffusive thickening \cite{landau1987fluid} such as the one plotted in Fig.\ref{fig:vortheo}(b), $\delta \sim (\nu H/\sin(\alpha)V)^{1/2}$, and we finally obtain the prediction: 

\begin{equation}
    \omega_{\rm interface} \sim \left(\frac{V^{3}\sin(\alpha)}{\nu H(\alpha)}\right)^{1/2}.
\label{eq:final}
\end{equation}

We compare this prediction with the measured $\omega_{\rm interface}$ for both sets of experiments (with and without dye in the jet) in Fig.\ref{fig:vortheo}(c) and obtain a satisfactory agreement, for a $\rm St = 0.013$, lower than what is reported for planar free shear layer. This discrepancy is attributed to the geometry of the shear layer, which is not planar in our system, as the flow separation is asymmetric around the jet. Nonetheless, the agreement between the predicted and measured angular frequency validates the proposed mechanism: to leading order, the wave field is forced by vortices emitted by the destabilization of the free shear layer. It is worth noting that ultimately, the development of the free shear layer is rooted in the asymmetric flow separation.

We now characterize the bubble cloud created by the cavity by measuring its bubble size distribution. Our experiments are conducted at high frame rates, which is essential to avoid counting the same bubbles multiple times. To do so, we identify the bubbles with a watershed algorithm, using Python and the skimage library  \cite{van2014scikit}. If an identified region has a circularity (parameter equal to $4\pi A/ P^2$ with $A$ the area of the region and $P$ its perimeter) smaller than 0.75, we consider the region to be an overlap of two bubbles. It is thus reanalyzed, and a distance transform is computed to obtain the markers for a second watershed. This method allows us to separate most of the overlaps between bubbles, as can be seen in the inset of Fig.\ref{fig:bubble}(a). The position and area of the bubble are then stored in a dataframe. We use trackpy  \cite{trackpy} to reconstruct the trajectories of the bubbles. Only the trajectories present on more than 10 images are kept, as a smaller trajectory is very likely to be a part of a longer trajectory that wasn't reconstructed fully. The mean radius of the bubble $R_b$ along this trajectory is measured, and we obtain the bubble size distribution for half a second of experiment, shown in Fig.\ref{fig:bubble}(a).

The distributions are asymmetric, with most of the bubbles smaller than the average radius of the bubble $\langle R_b\rangle$, and a few larger bubbles.

However, when bubble size distributions are rescaled by the total number of bubbles produced $N_{\rm tot}$ and their average bubble size, they all collapse, as seen in Fig.\ref{fig:bubble}(b). 
This collapse is less distinguishable for $R_b < \langle R_b \rangle$, as bubble detection is limited by the optical resolution $(\sim 3-10\ \rm \mu m$ depending on the experiment). 
The distribution can then be characterized with $\langle R_b\rangle$ and the total number of bubbles produced $N_{\rm tot}$. 

In this scenario of wave-induced bubble trapping, a correlation between the typical size of the wave $\lambda =2\pi/k$ and the size of the bubble is expected. 
In Fig.\ref{fig:bubble}(c), we can see that the average bubble size correlates with the wavelength, and that $\langle R_b\rangle \simeq 0.1\lambda$. 
One could also expect a correlation between the frequency of the waves and the total number of bubbles, but such correlation was not found in our experiments. 
This lack of direct frequency correlation implies that while the shear instability seeds the interface deformation, the ultimate pinch-off depends on other variables.

In summary, we have investigated bubble creation by the cavity formed by an inclined jet. 
We have shown that due to flow separation in the meniscus, a free shear layer appears and can destabilize into vortices. 
These vortices then interact with the interface of the cavity, forcing the frequency of the wavefield, whose wavelength is fixed by a shear-flow dispersion relation.
We propose a simple model for the selected frequency of the free shear layer, which is able to predict reasonably well the measured frequency of the cavity's wavefield. Finally, we measured the bubble size distribution in the bubble cloud produced by the cavity, and revealed a correlation between its average bubble size and the perturbation wavelength. 
These results constitute a first mechanistic description of air entrainment by an oblique water jet.

Several refinements could improve the description of this system. 
In particular, a criterion for wave destabilization, and a study of the ultimate destabilization of the waves would connect this work with recent advances on the break-up of gas filaments in turbulent flows  \cite{ni2024deformation,riviere2021sub,riviere2022capillary}.

\bibliography{jfm}

@PREAMBLE{
 "\providecommand{\noopsort}[1]{}" 
 # "\providecommand{\singleletter}[1]{#1}%" 
}

@article{ho1984perturbed,
  title={Perturbed free shear layers},
  author={Ho, C-M and Huerre, Patrick},
  journal={Annual review of fluid mechanics},
  volume={16},
  pages={365--424},
  year={1984}
}

@article{riviere2022capillary,
  title={Capillary driven fragmentation of large gas bubbles in turbulence},
  author={Rivi{\`e}re, Ali{\'e}nor and Ruth, Daniel J and Mostert, Wouter and Deike, Luc and Perrard, St{\'e}phane},
  journal={Physical Review Fluids},
  volume={7},
  number={8},
  pages={083602},
  year={2022},
  publisher={APS}
}

@article{guyot2020penetration,
  title={Penetration depth of a plunging jet: from microjets to cascades},
  author={Guyot, Gr{\'e}gory and Cartellier, Alain and Matas, Jean-Philippe},
  journal={Physical Review Letters},
  volume={124},
  number={19},
  pages={194503},
  year={2020},
  publisher={APS}
}

@article{riviere2021sub,
  title={Sub-Hinze scale bubble production in turbulent bubble break-up},
  author={Riviere, Ali{\'e}nor and Mostert, Wouter and Perrard, St{\'e}phane and Deike, Luc},
  journal={Journal of Fluid Mechanics},
  volume={917},
  pages={A40},
  year={2021},
  publisher={Cambridge University Press}
}

@article{ni2024deformation,
  title={Deformation and breakup of bubbles and drops in turbulence},
  author={Ni, Rui},
  journal={Annual Review of Fluid Mechanics},
  volume={56},
  number={1},
  pages={319--347},
  year={2024},
  publisher={Annual Reviews}
}

@article{trackpy,
author={Allan, D. B. and Caswell, T. and Keim, N. C. and van der Wel, C. M. and  Verweij, R. W},
    doi={https://doi.org/10.5281/zenodo.4682814},
    title={Trackpy v0.5.0},
    journal ={Zenodo},
    year={2021},
}

@article{koga1982bubble,
  title={Bubble entrainment in breaking wind waves},
  author={Koga, Momoki},
  journal={Tellus},
  volume={34},
  number={5},
  pages={481--489},
  year={1982},
  publisher={Wiley Online Library}
}

@article{detsch1990critical,
  title={The critical angle for gas bubble entrainment by plunging liquid jets},
  author={Detsch, Richard M and Sharma, Rajendra N},
  journal={The Chemical Engineering Journal},
  volume={44},
  number={3},
  pages={157--166},
  year={1990},
  publisher={Elsevier}
}

@article{van2014scikit,
  title={scikit-image: image processing in Python},
  author={Van der Walt, Stefan and Sch{\"o}nberger, Johannes L and Nunez-Iglesias, Juan and Boulogne, Fran{\c{c}}ois and Warner, Joshua D and Yager, Neil and Gouillart, Emmanuelle and Yu, Tony},
  journal={PeerJ},
  volume={2},
  pages={e453},
  year={2014},
  publisher={PeerJ Inc.}
}

@article{rabaud2020kelvin,
  title={The Kelvin--Helmholtz instability, a useful model for wind-wave generation?},
  author={Rabaud, Marc and Moisy, Fr{\'e}d{\'e}ric},
  journal={Comptes Rendus. M{\'e}canique},
  volume={348},
  number={6-7},
  pages={489--500},
  year={2020}
}

@misc{gaichies2026shapeinterfacehitoblique,
      title={Shape of an interface hit by an oblique jet}, 
      author={Theophile Gaichies and Anniina Salonen and Arnaud Antkowiak and Emmanuelle Rio},
      year={2026},
      eprint={2604.12788},
      archivePrefix={arXiv},
      primaryClass={physics.flu-dyn},
      url={https://arxiv.org/abs/2604.12788}, 
}

@misc{suppmat,
      title={See Supplemental Material at [URL]] for additional plots of the experimental data with color encoding different parameters and movies of the observed phenomena.}, 
}

@article{Mostert2022,
	author = {Mostert, W. and Popinet, S. and Deike, L.},
	journal = {J. Fluid Mech.},
	month = may,
	title = {High-resolution direct simulation of deep water breaking waves: transition to turbulence, bubbles and droplets production},
	volume = {942},
	year = {2022}}

@article{redor2025air,
  title={Air entrainment by turbulent plunging jets: Effect of jet roughness revisited},
  author={Redor, Ivan and Guyot, Gregory and Obligado, Martin and Matas, Jean-Philippe and Cartellier, Alain},
  journal={Chemical Engineering Science},
  volume={316},
  pages={121754},
  year={2025},
  publisher={Elsevier}
}

@book{landau1987fluid,
  title={Fluid Mechanics: Volume 6},
  author={Landau, Lev Davidovich and Lifshitz, Evgenii Mikhailovich},
  volume={6},
  year={1987},
  publisher={Elsevier}
}

@article{popinet2009accurate,
  title={An accurate adaptive solver for surface-tension-driven interfacial flows},
  author={Popinet, St{\'e}phane},
  journal={Journal of Computational Physics},
  volume={228},
  number={16},
  pages={5838--5866},
  year={2009},
  publisher={Elsevier}
}

@article{jeong1992free,
  title={Free-surface cusps associated with flow at low Reynolds number},
  author={Jeong, Jae-Tack and Moffatt, HK},
  journal={Journal of fluid mechanics},
  volume={241},
  pages={1--22},
  year={1992},
  publisher={Cambridge University Press}
}

@article{kiger2012air,
  title={Air-entrainment mechanisms in plunging jets and breaking waves},
  author={Kiger, Kenneth T and Duncan, James H},
  journal={Annual Review of Fluid Mechanics},
  volume={44},
  number={1},
  pages={563--596},
  year={2012},
  publisher={Annual Reviews}
}

@article{eggers2001air,
  title={Air entrainment through free-surface cusps},
  author={Eggers, Jens},
  journal={Physical review letters},
  volume={86},
  number={19},
  pages={4290},
  year={2001},
  publisher={APS}
}

@article{lorenceau2004air,
  title={Air entrainment by a viscous jet plunging into a bath},
  author={Lorenceau, {\'E}lise and Qu{\'e}r{\'e}, David and Eggers, Jens},
  journal={Physical review letters},
  volume={93},
  number={25},
  pages={254501},
  year={2004},
  publisher={APS}
}

@article{lorenceau2003fracture,
  title={Fracture of a viscous liquid},
  author={Lorenceau, Elise and Restagno, Fr{\'e}d{\'e}ric and Qu{\'e}r{\'e}, David},
  journal={Physical review letters},
  volume={90},
  number={18},
  pages={184501},
  year={2003},
  publisher={APS}
}

@article{bin1993gas,
  title={Gas entrainment by plunging liquid jets},
  author={Bi{\'n}, Andrzej K},
  journal={Chemical Engineering Science},
  volume={48},
  number={21},
  pages={3585--3630},
  year={1993},
  publisher={Elsevier}
}

@article{mckeogh1981air,
  title={Air entrainment rate and diffusion pattern of plunging liquid jets},
  author={McKeogh, EJ and Ervine, DA},
  journal={Chemical Engineering Science},
  volume={36},
  number={7},
  pages={1161--1172},
  year={1981},
  publisher={Elsevier}
}

@article{zhu2000mechanism,
  title={On the mechanism of air entrainment by liquid jets at a free surface},
  author={Zhu, Yonggang and O{\u{g}}uz, Hasan N and Prosperetti, Andrea},
  journal={Journal of Fluid Mechanics},
  volume={404},
  pages={151--177},
  year={2000},
  publisher={Cambridge University Press}
}

@article{miwa2019experimental,
  title={Experimental study of air entrainment rates due to inclined liquid jets},
  author={Miwa, Shuichiro and Xiao, Yi Geng and Saito, Yuya and Hibiki, Takashi},
  journal={Chemical Engineering \& Technology},
  volume={42},
  number={5},
  pages={1059--1069},
  year={2019},
  publisher={Wiley Online Library}
}

@article{chirichella2002incipient,
  title={Incipient air entrainment in a translating axisymmetric plunging laminar jet},
  author={Chirichella, D and Gomez Ledesma, R and Kiger, KT and Duncan, JH},
  journal={Physics of Fluids},
  volume={14},
  number={2},
  pages={781--790},
  year={2002},
  publisher={American Institute of Physics}
}
\end{document}